\documentclass[conference]{IEEEtran}
\IEEEoverridecommandlockouts
\usepackage{cite}
\usepackage{amsmath,amssymb,amsfonts}
\usepackage{algorithmic}
\usepackage{graphicx}
\usepackage{subcaption}
\usepackage{textcomp}
\usepackage[table]{xcolor}
\usepackage{braket}
\usepackage{array}
\usepackage{makecell}
\usepackage{graphicx}
\usepackage{booktabs}
\usepackage{listings}
\usepackage{xparse}
\usepackage{comment}
\newcommand{\ie}{\textit{i.e.}}
\newcommand{\viz}{\textit{viz.}}
\newcolumntype{s}{>{\columncolor{blue!30} \bf} c}
\newcolumntype{b}{>{\bf} c}
\def\BibTeX{{\rm B\kern-.05em{\sc i\kern-.025em b}\kern-.08em
    T\kern-.1667em\lower.7ex\hbox{E}\kern-.125emX}}

\begin{document}

\title{Transfer Learning Based Hybrid Quantum Neural Network Model for Surface Anomaly Detection\\
}

\author{\IEEEauthorblockN{Sounak Bhowmik}
\IEEEauthorblockA{\textit{Dept. of Electrical Engineering and Computer Science} \\
\textit{University of Tennessee, Knoxville}\\
Tennessee, USA \\
sbhowmi2@vols.utk.edu}
\and
\IEEEauthorblockN{Himanshu Thapliyal}
\IEEEauthorblockA{\textit{Dept. of Electrical Engineering and Computer Science} \\
\textit{University of Tennessee, Knoxville}\\
Tennessee, USA \\
hthapliyal@utk.edu}
}

\maketitle

\begin{abstract}
The rapid increase in the volume of data increased the size and complexity of the deep learning models. These models are now more resource-intensive and time-consuming for training than ever. This paper presents a quantum transfer learning (QTL) based approach to significantly reduce the number of parameters of the classical models without compromising their performance, sometimes even improving it. Reducing the number of parameters reduces overfitting problems and training time and increases the models' flexibility and speed of response. For illustration, we have selected a surface anomaly detection problem to show that we can replace the resource-intensive and less flexible anomaly detection system (ADS) with a quantum transfer learning-based hybrid model to address the frequent emergence of new anomalies better. We showed that we could reduce the total number of trainable parameters up to 90\% of the initial model without any drop in performance.
\end{abstract}


\section{Introduction}
Transfer learning~\cite{torrey2010transferLearning} is a truly biologically inspired approach, applying knowledge gained from a specific context to another. Quantum Transfer Learning (QTL)~\cite{wang2021quantum, mari2020transfer} is a branch of Quantum Machine Learning (QML) where we use the quantum mechanical properties offered by QML along with the knowledge transfer from benchmarked classical or quantum machine learning models. In this paper, we have applied QTL to reduce the number of parameters of a classical model, reducing its complexity without compromising its performance.  

\subsection{Motivation}
With the advancement of artificial intelligence (AI), AI models have significantly increased in complexity in the past few decades~\cite{menghani2023efficient}. We had a run for more sophisticated and powerful hardware to train bigger and more complex AI models. However, the more complex the model is, the more difficult it is to train, deploy, and maintain. Training a complex model is time-consuming and difficult because of factors like bias, variance, and local minima. Therefore, in this paper, we have tried to find out if reducing the number of parameters in a model without compromising its performance is possible. 

To demonstrate, we have picked a surface anomaly detection task and used quantum transfer learning (QTL) to approach this problem. Properties like superposition, entanglement, and interference bring parallelism and speed-up in a quantum computer. QML models are lightweight, flexible, and less resource-intensive than the bulky deep learning models used in industrial applications. Therefore, an efficient, lightweight hybrid quantum model can be considered a replacement for some portion of a bulky classical model to reduce its total trainable parameters and computational complexity, making them sustainable.

\subsection{Contributions of the paper}

In this paper, we proposed a classical-to-quantum transfer learning method for surface anomaly detection, which can be adopted easily in any other image-processing task. The contributions of this work are as follows,
\begin{itemize}
    \item We first trained three classical deep neural network models on a binary classification task based on the NEU-DET~\cite{song2013noiseNEU_DET} surface anomaly detection dataset. By random initialization, we ensured that the models did not suffer from problems like local optimum.
    \item We applied quantum transfer learning (QTL) to the classical models by replacing parts of the fully connected layers with a hybrid quantum neural network (dressed quantum network). We built three configurations of the QTL-based hybrid model for each classical model by substituting the last two, three, and all the dense layers. Therefore, we had nine hybrid models in total under this experiment.
    \item Keeping the initial block of classical layers fixed, we trained the replacement hybrid model through a 6-fold cross-validation process to assess their performance.
    \item From the recorded data, we compare the performance and the total number of parameters of the QTL-based hybrid models with their base classical model to show how they perform even after reducing a significant number of parameters. 
\end{itemize}

\subsection{Organization of the paper}
Section \ref{background} presents the background of quantum transfer learning. Next, in Section~\ref{workflow}, we described the workflow of building a quantum transfer learning-based model. In Section~\ref{proposed_model}, we discuss the structural details of the classical and the hybrid quantum models. Section \ref{experimental_setup} discusses the dataset features, problem formulation, training parameters, and hardware specifications used in the experiment. In Section \ref{results}, we show the simulation results of the transfer learning-based hybrid models and compare their performance and total number of parameters with the corresponding classical models. Finally, in Section \ref{discussion}, we summarize the insights we gain from the experimental results and the future expectations from quantum transfer learning.

\begin{figure}[htbp!]
    \centering
    \includegraphics[width=\columnwidth]{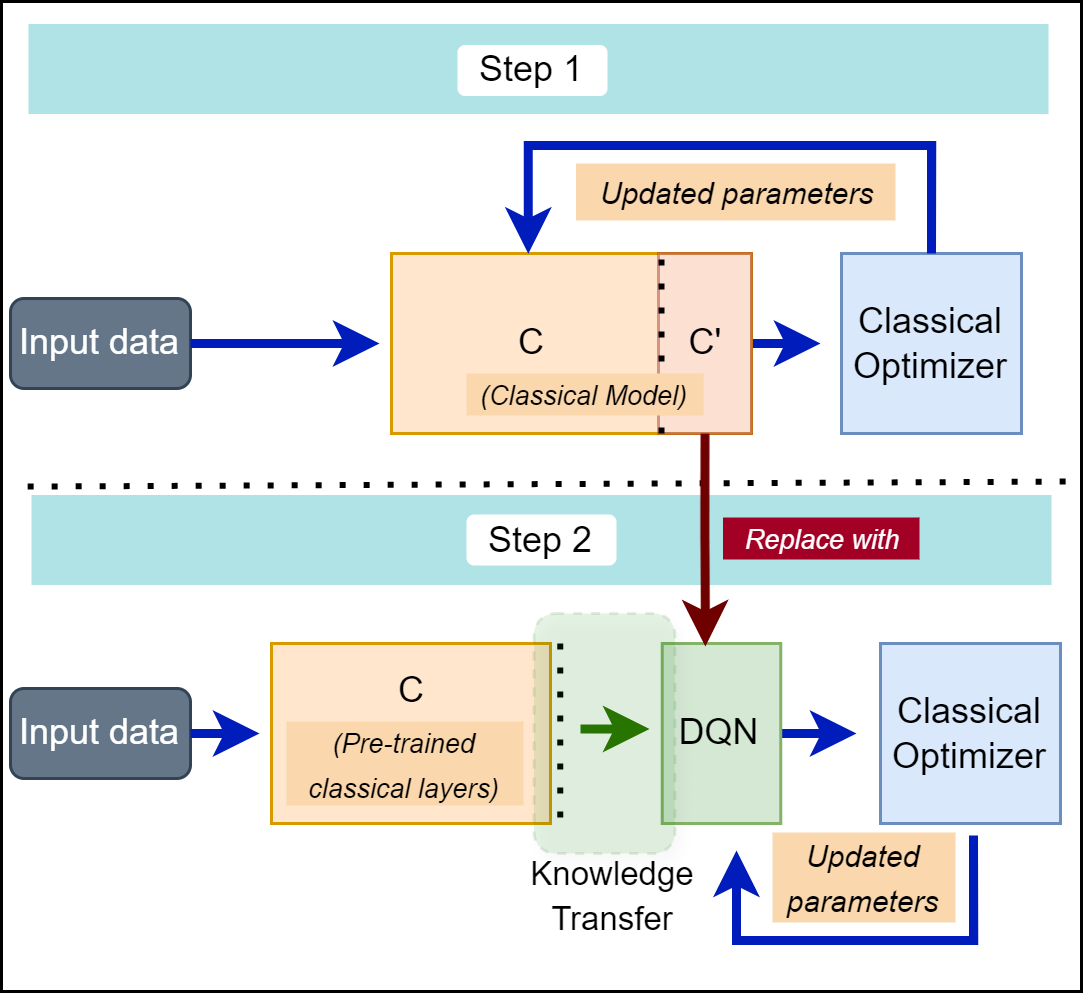}
    \caption{Procedure to build a quantum transfer learning based model.}
    \label{fig:TL_HQNN_WF}
\end{figure}

\begin{figure*}[htbp!]
    \centering
    \includegraphics[width=2\columnwidth]{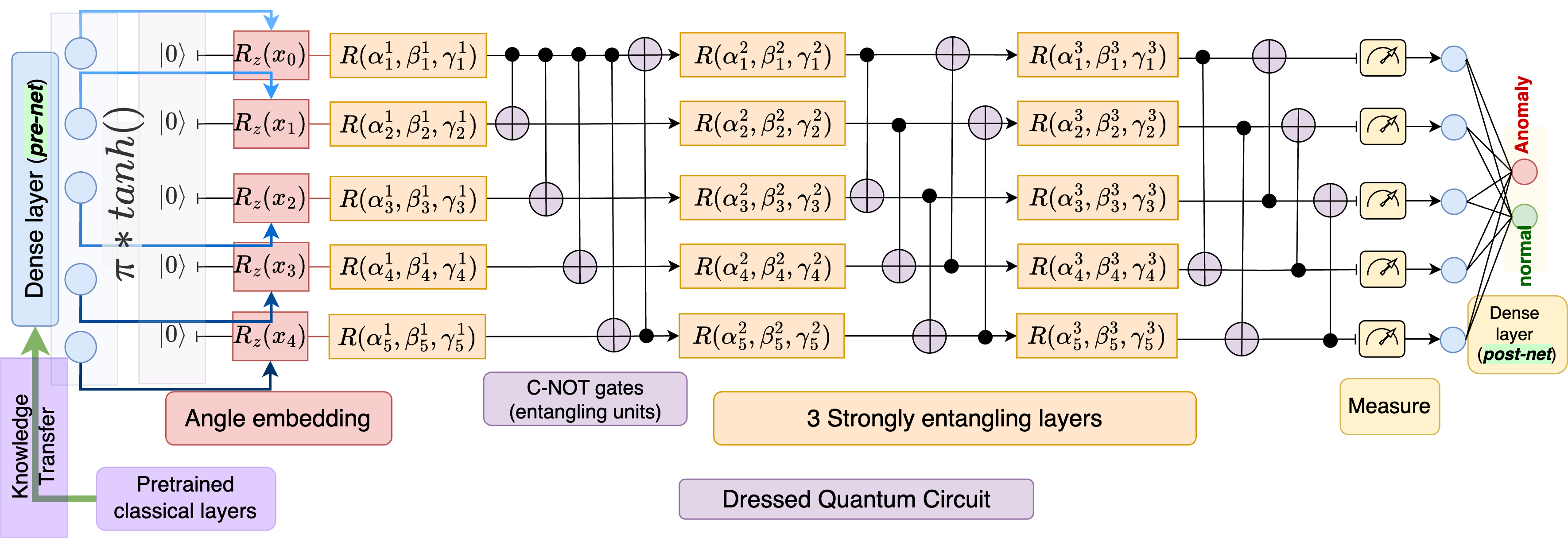}
    \caption{The structure of a dressed quantum network (DQN) consists of a variational quantum circuit (VQC) between a pre-processing dense layer (pre-net) and a post-processing dense layer (post-net).}
    \label{fig:TL_HQNN_SAD}
\end{figure*}
\section{Background} \label{background}
This section will discuss the concept of quantum transfer learning. First, we should look at a variational quantum circuit.

\subsection{Variational quantum circuit}
A variational quantum circuit, or VQC, is a perfect analogy to a classical deep neural network. It consists of multiple layers of several single qubit rotational gates, followed by an array of entangling layers. The rotational gates are controlled by classical parameters, which can be tuned by any classical optimization method, such as gradient descent. The entangling layers consist of controlled-NOT gates. 

A \textit{quantum layer} can be defined as a unitary operation ($U$), which is applied on the input quantum state $\ket{X_{n_q}}$, consisting of $n_q$ quantum subsystems or qubits, to produce an output state of $\ket{Y_{n_q}}$. 
\begin{equation}
    L : U(w)\ket{X_{n_q}} = \ket{Y_{n_q}}
\end{equation}
Here, $w$ is a set of classical trainable parameters. 
\begin{comment}
    Unlike the classical feed-forward layers, the quantum variational circuit preserves the input dimension of the quantum state, which matches the input dimension. A $d$-layered VQC can be visualized as,
\begin{equation}
    \Theta: L_d \circ L_{d-1} \circ ... L_1
\end{equation}
\end{comment}

To use a VQC to process classical features, we need to embed the classical features into the corresponding quantum states first. The embedding layer has a set of parameterized rotational gates, which, upon application on a basis state $\ket{0^{\otimes n_q}}$, controlled by the classical feature vector $x_{n_q}$, produces an embedded quantum state $\ket{x^{\otimes n_q}}$.
\begin{equation}
    \mathcal{E}: U(x_{n_q})\ket{0^{\otimes n_q}} = \ket{x^{\otimes n_q}}
\end{equation}
However, to get the classical output vector $y$, we need to measure the expectation values of $n_q$ observables, $\hat{z} = [\hat{z_1}\hat{z_2}...\hat{z_{n_q}}]$, at the output. We can represent this measurement layer as,
\begin{equation}
    M: \ket{x^{\otimes n_q}} \rightarrow y = \braket{x^{\otimes n_q}|\hat{z}|x^{\otimes n_q}}
\end{equation}
Therefore, a VQC is a sequence of all these layers, viz.,
\begin{equation}
    \mathbb{Q} = M\circ L \circ \mathcal{E}
\end{equation}

\subsection{Quantum transfer learning}
Quantum transfer learning can be applied to a pre-trained classical deep learning model by replacing their last few dense layers with a VQC or a hybrid quantum model, where VQC has pre-processing and post-processing classical layers. 

Assume $C$ is a pre-trained classical model's initial set of layers, specialized in feature extraction. Then, the transfer learning-based quantum model can be built as,
\begin{equation}\label{trf_lr}
    T = L_{n_q \rightarrow n_o} \circ \mathbb{Q} \circ L_{n_c \rightarrow n_q}\circ C
\end{equation}
In Eqn.~\ref{trf_lr}, $\mathbb{Q}$ is a VQC consisting of $n_q$ qubits, $n_c$ is the dimension of the output vectors from $C$, and $n_o$ is the number of output classes in the given problem. $L_{n_c \rightarrow n_q}$ is a preprocessing classical dense layer, \textit{pre-net} producing output with a dimension of $n_q$. $L_{n_q \rightarrow n_o}$ is a postprocessing classical dense layer, \textit{post-net}, that outputs a probability distribution over all the classes. The use of classical layers brings flexibility to the design of the VQC. That means the number of qubits used in the VQC does not need to depend on the output dimension of the pre-trained network or the number of output classes. The set-up $L_{n_q \rightarrow n_o} \circ \mathbb{Q} \circ L_{n_c \rightarrow n_q}$ in Eqn.~\ref{trf_lr}, is called a dressed quantum network~\cite{mari2020transfer}.

The total number of parameters in a dressed quantum network can be calculated using Eqn.~\ref{eqn:dqn_params}. Here, $n_{ip}$ is the input feature dimension of the \textit{pre-net}, $n_{q}$ = the number of qubits in the VQC, $n_{d}$ = the number of fully entangled layers or the depth of the VQC, and $n_{c}$ = the number of output classes.

\begin{equation}\label{eqn:dqn_params}
\begin{split}
    W_{dqn} = W_{pre-net} +W_{VQC}  + W_{post-net}\\
    W_{pre-net} = n_{ip}*n_{q} + n_{q}\\
    W_{VQC} = 3*n_{q}*n_{d}\\
    W_{post-net} = n_{q}*n_{c} + n_{c}
\end{split}
\end{equation}

In the upcoming sections, we will discuss the architecture of our transfer learning-based setup and compare its performance against the corresponding LeNet-based \cite{LeNet} classical models.

\begin{table*}[htbp]
    \caption{Architectures of different classical convolutional models used in this experiment}
    \label{tab:arch_cm}
    \centering
    \resizebox{2.1\columnwidth}{!}{
    \setlength{\tabcolsep}{2pt}
    \begin{tabular}{|c|m{12cm}|s|} \hline 
         \rowcolor{white}\makecell{\textbf{Model Name}}&  \makecell{\textbf{Architecture}} & \makecell{\bf Number of \\ \bf Parameters}\\ \hline 
          Model-1&  Conv2D (1, 32, 32, 2) $\circ$ MaxPool2D (8, 1) $\circ$ Conv2D (32, 64, 16, 2) $\circ$ MaxPool2D (8, 1) $\circ$ Conv2D (64, 128, 16, 2) $\circ$ MaxPool2D (8, 1) $\circ$ Conv2D (128, 128, 2, 1)  $\circ$ MaxPool2D (8, 2) $\circ$ Flatten () $\circ$ Linear (3200, 128) $\circ$ dropout (p=0.5) $\circ$ Linear (128, 64) $\circ$ dropout (p=0.25) $\circ$ Linear (64, 32) $\circ$ dropout (p=0.12) $\circ$ Linear (32, 16) $\circ$ Linear (16, 2)& 1,076,338\\ \hline
          
         Model-2&  Conv2D (1, 32, 4, 2) $\circ$ MaxPool2D (4, 2) $\circ$ Conv2D (32, 64, 8, 2) $\circ$ MaxPool2D (2, 2) $\circ$ Conv2D (64, 128, 4, 2) $\circ$ Flatten () $\circ$ Linear (2048, 128) $\circ$ dropout (p=0.5) $\circ$ Linear (128, 64) $\circ$ dropout (p=0.25) $\circ$ Linear (64, 16) $\circ$ Linear (16, 2)& 534,482\\ \hline 
         
         Model-3&  Conv2D (1, 32, 8, 2) $\circ$ MaxPool2D (4, 2) $\circ$ Conv2D (32, 64, 4, 2) $\circ$ MaxPool2D (4, 1) $\circ$ Conv2D (64, 128, 2, 1) $\circ$ MaxPool2D (4, 2) $\circ$ Flatten () $\circ$ Linear (8192, 128) $\circ$ dropout (p=0.5) $\circ$ Linear (128, 64) $\circ$ dropout (p=0.25) $\circ$ Linear (64, 16) $\circ$ Linear (16, 2)& 1,125,842\\ 
         \hline 
         \bottomrule
    \end{tabular} 
    }
    \footnotesize{---\\The general structure of the layers is as follows: Conv2D (input channels, output channels, kernel size, stride), MaxPool2D (kernel size, stride), Linear (input feature dimension, output feature dimension), dropout (p = the probability of an element being set to zero).}
\end{table*}

\section{Workflow} \label{workflow}
There are two major steps in building a quantum transfer learning-based model.
\begin{itemize}
    \item We first need to train a classical model. For that, we select a dataset and train a classical model until we get a reasonable performance. In general, for image-processing applications, the models consist of an initial block of convolution and pooling layers responsible for feature extraction, followed by fully connected dense layers that classify the data based on the extracted features. In step 1 of Fig.~\ref{fig:TL_HQNN_WF}, these two blocks have been identified as C and C'. 

    \item In the second step (refer to step 2 of Fig.~\ref{fig:TL_HQNN_WF}), we replace the second block of classical layers, C', with a dressed quantum circuit (DQN). To build the DQN, we chose the architecture of the variational quantum circuit (VQC) as we prefer. There are three primary parameters to determine its structure, \viz, the number of qubits, the type of entangling layers, and the number of layers. Depending on the requirements and available resources, we can choose different configurations. For more information on the structure of VQC, refer to \cite{sim2019expressibility}. Finally, we need two more classical dense layers for pre-processing and post-processing data (pre-net and post-net, respectively). Refer to Fig.~\ref{fig:TL_HQNN_SAD} for the detailed structure of the DQN we used in our experiment.
\end{itemize}
In the second step, we only update the parameters in the DQN, keeping the classical layers fixed at their pre-trained weights. This process ensures that we get the superior features extracted by the benchmark classical layers and then leverage the quantum mechanical properties of the DQN for feature fusion and classification. As we replace the bulky, dense layers with a relatively simple hybrid quantum-classical architecture, we can significantly reduce the total number of parameters of the working model. The parameter reduction can be calculated using Eqn.~\ref{eqn:param_reduc}, where $W_i$ is the total number of parameters in $i$th classical layer and $x$ is the number of replaced layers. $W_{dqn}$ is the number of parameters in the dressed quantum circuit, calculated using Eqn.~\ref{eqn:dqn_params}.
\begin{equation} \label{eqn:param_reduc}
    R = \frac{\sum_{i=1}^{x}{W_i}-W_{dqn}}{\sum_{i=1}^{x}{W_i}}*100\%
\end{equation}

\section{Proposed Model} \label{proposed_model}
In this experiment, we have built three classical models using LeNet-based architecture. The details, such as the number and the dimension of the filters in the convolution layers, the total number of layers, and the number of units per layer, can be found in Table~\ref{tab:arch_cm}. 

We also built three different configurations of quantum transfer learning-based models based on each classical model. We sequentially replaced the fully connected dense layers in our model, starting with two layers, followed by three, and finally, all of them, using dressed quantum circuits in each case. Refer to Fig. \ref{fig:TL_HQNN_SAD} for the detailed architecture of the VQC, which consists of three parts. 
\begin{itemize}
    \item The first part, consisting of an array of single-qubit rotational gates, embeds the classical features extracted through the classical layers into quantum states. The rotational gates are controlled by the output of the \textit{pre-net}, meaning that the rotation will depend on the classical features, and the output will be a quantum map of them.
    
    \item Next is a 3-layered, strongly entangled quantum circuit (VQC) comprising five qubits. These layers comprise a set of parameterized rotational gates and an array of C-NOT gates that entangle the qubits. We have kept a low number of qubits and layers as the complexity rises exponentially with increasing depth. Fig. \ref{fig:TL_HQNN_SAD} shows the detailed architecture of the VQC.

    \item Ultimately, we measure the output in the $Z$-basis and get an expectation value used by the \textit{post-net}.
\end{itemize}

The pre-net takes the extracted features from the pre-trained classical layers as an input, and its output dimension matches the number of qubits of the VQC. The dimension of the output of the pre-net must match the number of qubits in the circuit for quantum embedding to work. The output of the post-net gives us the probability distribution over two classes: anomalous and normal. There are 45 trainable classical parameters in the VQC of every QTL-based hybrid model. However, the number of parameters in \textit{pre-net} varies according to the number of features extracted from the previous layers.

%=========================================================================
% Types of defects
\begin{figure}[ht!]
    \centering
    \begin{subfigure}[b]{0.3\columnwidth}
        \includegraphics[width=\linewidth]{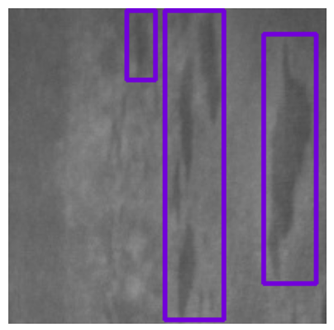}
        \caption{Inclusion}
    \end{subfigure}
    \hfill % This command adds space between the figures horizontally.
    \begin{subfigure}[b]{0.3\columnwidth}
        \includegraphics[width=\linewidth]{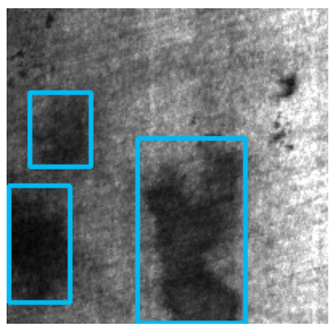}
        \caption{Patches}
    \end{subfigure}
    \hfill % This command adds space between the figures horizontally.
    \begin{subfigure}[b]{0.3\columnwidth}
        \includegraphics[width=\linewidth]{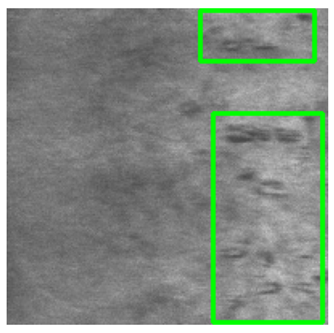}
        \caption{Rolled-in scale}
    \end{subfigure}
 
    \begin{subfigure}[b]{0.3\columnwidth}
        \includegraphics[width=\linewidth]{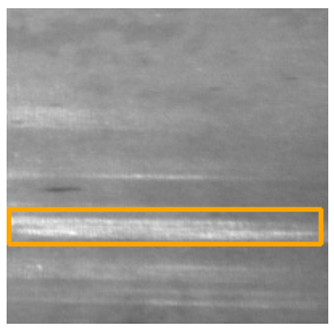}
        \caption{Scratches}
    \end{subfigure}
    \hfill
    \begin{subfigure}[b]{0.3\columnwidth}
        \includegraphics[width=\linewidth]{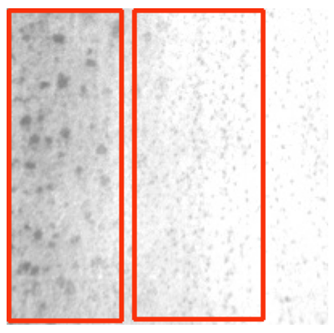}
        \caption{Pitted surface}
    \end{subfigure}
    \hfill
    \begin{subfigure}[b]{0.3\columnwidth}
        \includegraphics[width=\linewidth]{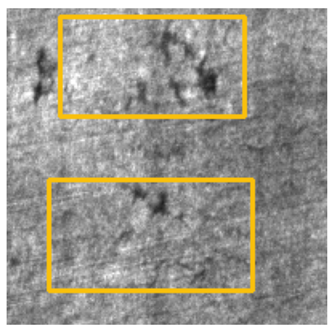}
        \caption{Crazing}
    \end{subfigure}
 
    \caption{Different types of surface anomalies in NEU-DET.}
    \label{fig:neu_det_defect_types}
\end{figure}
%=========================================================================

%=========================================================================
% Hyperparameters Table
\begin{table}[htbp]
\caption{Training hyperparameters and hardware specs}
    \label{tab:training_specs}
    \centering
    \resizebox{0.75\columnwidth}{!}{
    \setlength{\tabcolsep}{8pt}
    \begin{tabular}{|l|c|}
    \hline
         \rowcolor{white}\bf{Parameter}& \bf{Value}\\
         \hline
         \rowcolor{blue!30}\bf Common& \\
         Number of classes& 2\\
         Batch size& 64 \\
         Input image size& 1x200x200\\
         Optimiser & Adam\\
         Loss function& Cross-entropy loss\\
         Processor& i7-10700 CPU @ 2.90GHz\\
         GPU & NVIDIA Tesla T4\\
         RAM & 51 GB\\
         
         \rowcolor{blue!30}\bf Classical& \\
         Learning rate& 0.001\\
         Activation functions & \makecell{soft-max (output), ReLU(others)}\\
         Epochs& 120 \\
         Framework & Pytorch\\
         
         \rowcolor{blue!30}\bf Hybrid quantum& \\
         Number of qubits& 5\\
         Number of layers& 3\\
         Type  of layers& Strongly entangled\\
         Measurement & Z-basis\\
         Learning rate& 0.0008\\
         Activation functions & \makecell{tanh (pre-net), soft-max (post-net)\\ReLU(VQC)}\\
         Epochs& 40 \\
         Framework & Pennylane\\
         Simulator & \makecell{default.qubit, lightning.gpu}\\
         \hline
         \bottomrule
    \end{tabular}
    }
\end{table}
%=========================================================================

\section{Experimental Set up} \label{experimental_setup}
This section will discuss the used dataset, problem formulation, and training specification for the classical and hybrid working models. 

\subsection{Dataset (pre-processing and clean-up steps)}
For this experiment, we have used the NEU-DET dataset, which has 300 images of each of the following six types of anomalies of the steel surface: \textit{inclusion, patches, rolled-in scale, scratches, pitted surface, crazing}. Each image has a corresponding annotation file containing the location of the bounding box around the anomaly. Examples of each type of defect are shown in Fig. \ref{fig:neu_det_defect_types}. All of the images are square-shaped RGB images of dimension 200x200. We initially made them gray-scale and standardized their value.

\subsection{Problem formulation}
In this work, we have formulated a binary classification problem, i.e., our classifiers can differentiate between an anomalous image and a normal one. As we did not have any normal images in the dataset, we had to generate a repository of normal surface patches by cropping the defect-free regions of the available images and resizing them to the original image dimension, 200x200. However, we saw many \textit{pitted surface} and \textit{crazing} type images had defects outside the annotated regions. So, we dropped these two types of defects. After clean-up, we had 1103 anomalous images, and we randomly selected 1103 normal images from our previously generated repository, which produced a total of 2206 images in our customized dataset.

\subsection{Training specification}
We initially trained three classical models on this data for 120 epochs, with a test size 20\%. We randomly initialized each model 5 times to avoid poor local minima problems. From them, we selected the best-performing model from each architecture. So, after this exhaustive search, we found three different models on which we apply quantum transfer learning in the later stage. 

Then, we derived three QTL-based hybrid models from each classical architecture in the next step. Then, we performed 6-fold cross-validation for each of them. We recorded the results after 40 epochs for each fold. We have used a 5-qubit, 3-layered, strongly entangled VQC and Pennylane's \texttt{default.qubit} and \texttt{lightning.gpu} simulators to run the simulations.
For a complete list of the training hyperparameters and the hardware specifications, refer to Table \ref{tab:training_specs}.

We kept the number of qubits in the VQC at 5 to avoid resource and time constraints. However, increasing the number of qubits in the circuit will benefit feature extraction or fusion tasks.

%================================================================================
\begin{table}[htbp]
\caption{Performance and parameters comparison between the classical and their corresponding quantum transfer-learning-based models}
    \label{tab:performance_parameter_comparison}
    \centering
    \resizebox{\columnwidth}{!}{
    \setlength{\tabcolsep}{8pt}
    \begin{tabular}{|l|c|c|c|c|}
    \hline
         \rowcolor{white}\bf{Model}& \bf{F1 score (\%)}&\makecell{\bf{Boost}\\ \bf{Perf (\%)}}& \makecell{\bf{\#Params}}&  \makecell{\bf{Reduc.}\\ \bf{\#Params (\%)}} \\
         \hline
         \rowcolor{blue!30}CM-1& 88.58&-&	1,076,338& -\\
         %\makecell[l]{QTL-M-1}& 90.38&	2.03&   1,076,446&		-0.01\\
         \makecell[l]{QTL-M-1}& 88.91&	0.38&   1,074,078&		0.21\\
         \makecell[l]{QTL-M-2}& 88.62&	0.05&   1,066,142&		0.95\\
         \makecell[l]{QTL-M-3}& 90.08&	1.69&   671,764&        37.59\\
         \hline
         \rowcolor{blue!30}CM-2& 95.16    & -	&534,482&-\\
         %\makecell[l]{QTL-M-1}& 96.54&	1.45&   534,590&		-0.02\\
         \makecell[l]{QTL-M-1}& 96.62&	1.54&   533,790&		0.13\\
         \makecell[l]{QTL-M-2}& 96.81&	1.73&   525,824&		1.62\\
         \makecell[l]{QTL-M-3}& 98.95&	3.98&   273,182&		48.89\\
         \hline
         \rowcolor{blue!30}CM-3& 97.32&  -&	1,125,842& -\\
         %\makecell[l]{QTL-M-1}& 97.55&	0.24&   1,125,950&		-0.01\\
         \makecell[l]{QTL-M-1}& 97.81&	0.50&   1,125,150&		0.06\\
         \makecell[l]{QTL-M-2}& 97.90&	0.60&   1,117,214&		0.77\\
         \makecell[l]{QTL-M-3}& 97.85&	0.55&   108,830&	    	90.33\\
         \hline
         \bottomrule
    \end{tabular}
    }
    \footnotesize{ \break \\ The information of the three configurations of quantum transfer learning-based models based on each classical model is represented as the CM 1 to 3 and QTL-M 1 to 3.}
\end{table}

\section{Results} \label{results}

\begin{figure}[htbp]
    \centering
    \includegraphics[width=0.9\linewidth]{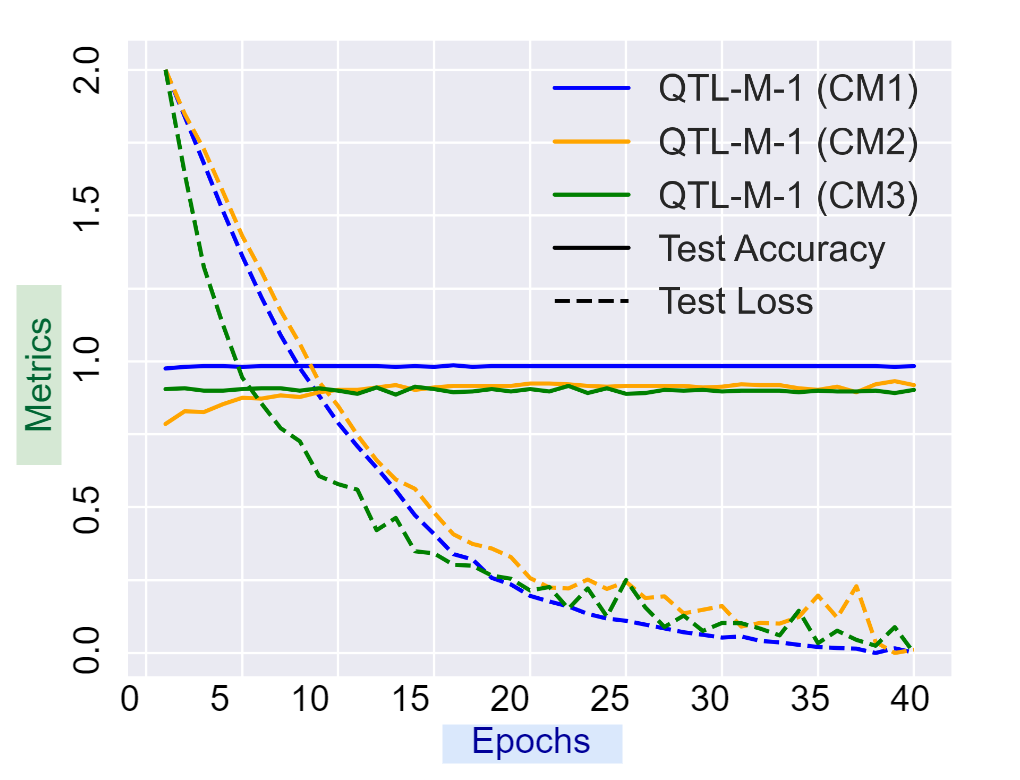}
    \caption{Convergence plot: Transfer learning-based models, with the last \textit{two} dense layers substituted by the dressed quantum network.}
    \label{fig:convergence_hybrid_2}
\end{figure}

\begin{figure}[htbp]
    \centering
    \includegraphics[width=0.9\linewidth]{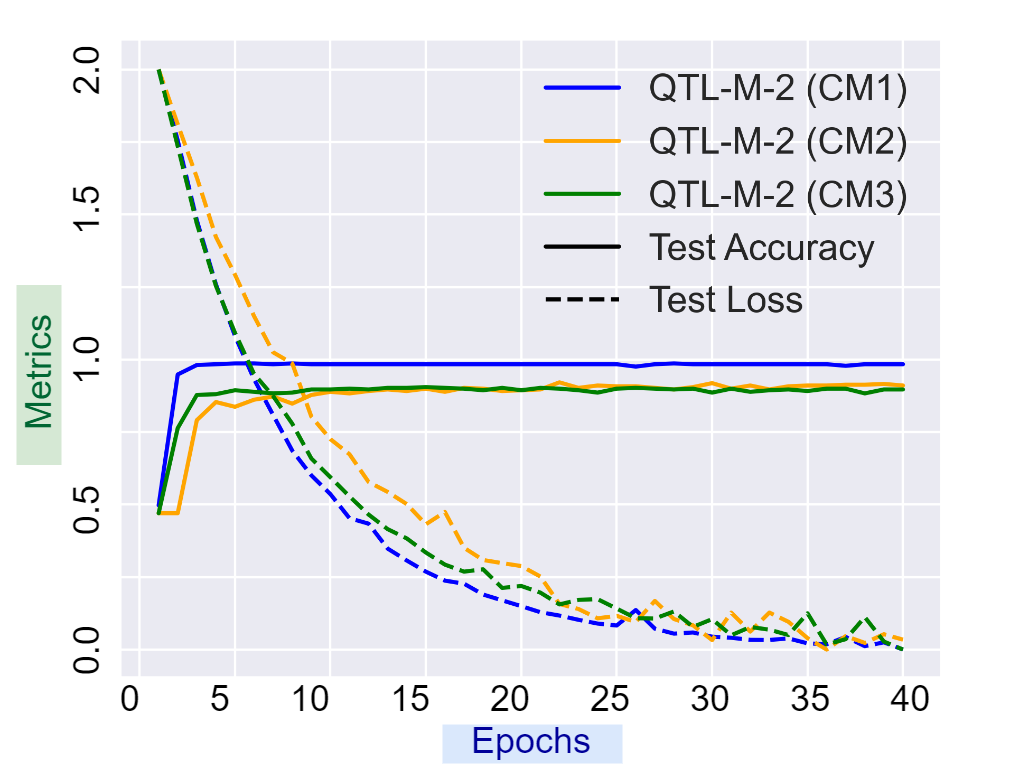}
    \caption{Convergence plot: Transfer learning-based models, with the last \textit{three} dense layers substituted by the dressed quantum network.}
    \label{fig:convergence_hybrid_3}
\end{figure}

\begin{figure}[htbp]
    \centering
    \includegraphics[width=0.9\linewidth]{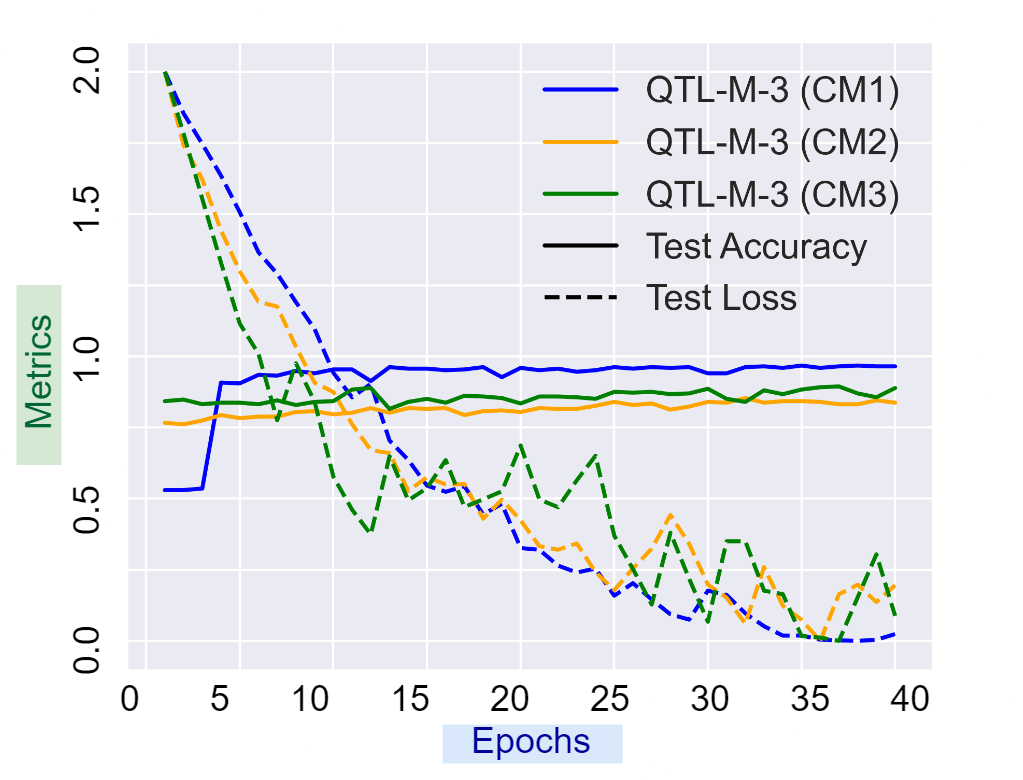}
    \caption{Convergence plot: Transfer learning-based models, with \textit{all but the convolutional} layers substituted with the dressed quantum network.}
    \label{fig:convergence_hybrid_all}
\end{figure}

We have trained and validated the three classical models using the same training and validation dataset. While training the classical models, we recorded the test accuracy after every epoch. On top of each of the classical models, we built three quantum transfer learning-based models and trained them through a 6-fold cross-validation process. We have shown the convergence of these models in Figures  \ref{fig:convergence_hybrid_2}, \ref{fig:convergence_hybrid_3}, and \ref{fig:convergence_hybrid_all}. We have normalized the test losses between 0 and 2 for better visualization. In Table~\ref{tab:performance_parameter_comparison}, we have listed the F1 score of the classical models, the average F1 score of the QTL-based models, and the total number of parameters of all individual models. 

The first classical model, CM-1, achieved an F1 score of 88.58\% on the binary classification problem. The third QTL-based model based on CM-1, QTL-M-3, got an average F1 score of 90.08\%, where we get a 1.7\% performance boost, and it has 37.59\% less parameter than CM-1.

The second classical model, CM-2, achieved an F1 score of 95.16\%. The third QTL-based model based on CM-2, QTL-M-3, got an average F1 score of 98.95\%, where we get an almost 4\% performance boost, and it has 49\% less parameter than CM-2. Even for the other QTL-based models, we achieved an increased performance. However, the total parameters were mostly the same in number.

The third classical model, CM-3, achieved an F1 score of 97.32\%. It was, in fact, the best-performing classical model. The third QTL-based model based on CM-3, QTL-M-3, got an average F1 score of 97.85\%. Even though the performance did not improve much, we could reduce the total number of parameters by 90\%. In the other QTL-based configurations, the performance did not decrease either.

The QTL-based hybrid configurations 1 and 2 did not appreciate much parameter reduction. There is a good reason for that. Following the block of convolution and pooling, the first dense layer comprises the maximum number of parameters. We replace this layer for all the classical models only in the QTL-M-3 configuration. That is why there is a considerable reduction in the number of parameters in the third configuration and a minimal reduction for the rest. 

Following Table~\ref{tab:performance_parameter_comparison}, we can see that, besides the reduction in the total number of parameters, the F1 scores of the quantum transfer learning-based models have also increased compared to the base classical model. Therefore, in a tradeoff between performance and model complexity, we saw that without any compromise in the performance, we could significantly reduce the complexity, \ie, the total number of parameters in a classical model using a quantum transfer learning-based hybrid approach.

%==========================================================================================
%pLOTS

%==========================================================================================
\section{Conclusion} \label{discussion}
In this experiment, we built three classical models. Then, we applied quantum transfer learning, replacing different blocks of fully connected layers with a hybrid quantum circuit. In this process, we could significantly reduce the total number of parameters in the models and improve their performance. We could achieve this by using the efficient feature extraction capability of the classical convolution layers and the quantum circuit's feature fusion and mapping capability due to their unique quantum mechanical properties. Since a VQC can be trained to behave approximately the same way as the fully optimized circuit, which certainly optimizes the objective function, we could use this behavior and replace the heavy classical layers with a simpler VQC that could approximate its job very well.

Reduction in the total number of parameters will reduce the complexity, making the models more sustainable and easy to train. Other benefits include reduced training time, faster inference, and easy deployment and maintenance. Lightweight quantum models are more flexible than similar classical models; therefore, we can easily modify them on demand.  

In the current NISQ era, we can best utilize the power of quantum machine learning models with efficient hybrid approaches. Because using a hybrid approach mitigates the hardware limitations, such as a limited number of qubits and noisy quantum subsystems. On top of that, we can leverage the power of quantum processing to empower the classical benchmark models.

\bibliographystyle{IEEEtran}
\bibliography{IEEEabrv}

\end{document}